# Nonlinear infrared spectroscopy free from spectral selection

Anna Paterova[1,3], Shaun Lung[1,2], Dmitry Kalashnikov[1], and Leonid A. Krivitsky[1*]

[1] Data Storage Institute, Agency for Science Technology and Research (A*STAR), 138634 Singapore

[2] Department of Physics, National University of Singapore, 117542, Singapore

[3] School of Electrical and Electronic Engineering, Nanyang Technological University, 639798, Singapore

* Leonid-K@dsi.a-star.edu.sg

**ABSTRACT**

Infrared (IR) spectroscopy is an indispensable tool for many practical applications including material analysis and sensing. Existing IR spectroscopy techniques face challenges related to the inferior performance and the high cost of IR-grade components. Here, we develop a new method, which allows studying properties of materials in the IR range using only visible light optics and detectors. It is based on the nonlinear interference of entangled photons, generated via Spontaneous Parametric Down Conversion (SPDC). In our interferometer, the phase of the signal photon in the visible range depends on the phase of an entangled IR photon. When the IR photon is traveling through the media, its properties can be found from observations of the visible photon. We directly acquire the SPDC signal with a visible range CCD camera and use a numerical algorithm to infer the absorption coefficient and the refraction index of the sample in the IR range. Our method does not require the use of a spectrometer and a slit, thus it allows achieving higher signal-to-noise ratio than the earlier developed method.

**INTRODUCTION**

Infrared (IR) spectroscopy is an established tool for material analysis, and it is widely adopted by a number of industries[1-6]. Historically, IR spectroscopy started with direct transmission techniques, which were later superseded by Fourier Transform IR (FTIR) spectroscopy[7]. Extensive efforts on the development of the FTIR spectroscopy for the last 40 years resulted in its wide adoption and successful commercialization. However, FTIR spectroscopy still faces technical challenges, including low efficiency and high dark noise of IR photodetectors (which often require cryogenic cooling), absorption of the signal by water vapor and the requirement of using specialized IR-grade optics[8].

The above-mentioned reasons make further development of traditional IR techniques challenging. At the same time, a number of techniques using entangled states of light have been suggested and demonstrated. They offered certain beneficial features and extended the functionality of traditional approaches[9-12]. Recently a simplified approach, which allows revealing the IR properties of the medium from the measurements of the visible light was demonstrated[13]. The concept is based on the nonlinear interference of entangled photons produced via spontaneous parametric down conversion (SPDC)[14,15]. The properties of the media in the IR range are found from the interference pattern of the visible photons. The method uses well developed and cost-efficient visible light components.

However, the technique requires spectral and spatial filtering of visible photons, performed by a spectrograph and a slit. The filtering reduces the signal throughput and adds complexity. In this work, we develop a method, which allows us to eliminate the need for the spectrograph and the slit. We achieve a higher signal-to-noise ratio, reduce the system footprint and make it more affordable. The technique is tested with an IR absorption line of the carbon dioxide gas at 4.3 microns, which is widely used for environmental sensing and biomedical applications.

**RESULTS**

**Theoretical framework**

The SPDC occurs in a medium with nonlinear susceptibility $\chi^{(2)}$, where the pump laser photon splits into a pair of highly correlated photons (referred to as signal and idler) known as biphoton[16,17]. The process yields conservation of energy and momentum, known as phase matching conditions:

$$\omega_p = \omega_s + \omega_i; \quad \vec{k}_p = \vec{k}_s + \vec{k}_i \qquad (1)$$

where $\omega_{p,s,i}$ are the frequencies and $\vec{k}_{p,s,i}$ are the wave vectors of the pump, signal and idler photons.

The biphoton state vector in the frequency-angular domain has the following form:

$$|\psi\rangle = \int d\omega_s d\vec{k}_s d\omega_i d\vec{k}_i \, F(\omega_s, \vec{k}_s, \omega_i, \vec{k}_i) a_s^+(\omega_s, \vec{k}_s) a_i^+(\omega_i, \vec{k}_i)|vac\rangle \qquad (2)$$

where $a_{s,i}^+$ are the photon creation operators for the signal and idler modes, respectively, $F(\omega_s, \vec{k}_s, \omega_i, \vec{k}_i)$ is the biphoton field amplitude, which modulus square defines the probability of the biphoton generation.

In the monochromatic and plane-wave pump approximation, the amplitude of the signal photon for a uniform crystal ($\chi^{(2)}$ =const) is given by the integration of $F(\omega_s, \vec{k}_s, \omega_i, \vec{k}_i)$ over all unregistered idler photons, resulting in the following[16,18]:

$$F_1(\omega_s, \theta_s) \propto sinc\left(\Delta k(\omega_s, \theta_s)\frac{L}{2}\right) \qquad (3)$$

$$\Delta k(\omega_s, \theta_s) = \vec{k}_p - \vec{k}_s - \vec{k}_i = \frac{n_p \omega_p}{c} - \frac{n_s \omega_s}{c}\sqrt{1 - \left(\frac{\sin\theta_s}{n_s}\right)^2} - \frac{n_i(\omega_p - \omega_s)}{c}\sqrt{1 - \left(\frac{\omega_s \sin\theta_s}{n_p(\omega_p - \omega_s)}\right)^2} \qquad (4)$$

where $\Delta\vec{k}$ is the wave vectors mismatch in the crystal, $\theta_s$ is the external angle between the pump and the signal wave vectors, $n_{p,s,i}$ are crystal refractive indices at the pump, signal and idler photon wavelengths, respectively, and $L$ is the thickness of the crystal. The typical angular-wavelength distribution of the biphoton amplitude $|F_1(\omega_s, \theta_s)|^2$ is shown in Fig. 1a.

Let us now consider the setup, where two equivalent crystals are placed sequentially into a common coherent pump with some distance $L'$ between them[13,18]. In this case, biphotons, generated in the first crystal, interfere with the ones generated in the second crystal. Assuming that all the photons are within the transverse interaction region, the interference pattern of the signal photons is given by[18-22]:

$$|F_2(\omega_s, \theta_s)|^2 \propto sinc^2\left(\Delta k(\omega_s, \theta_s)\frac{L}{2}\right)(1 + |\tau| \cdot \cos(\Delta\varphi)) \quad (5a)$$

$$\Delta\varphi \equiv \Delta k(\omega_s, \theta_s)L + \Delta k'(\omega_s, \theta_s)L' \quad (5b)$$

$$\Delta k'(\omega_s, \theta_s) = \vec{k}'_p - \vec{k}'_s - \vec{k}'_i = \frac{n'_p \omega_p}{c} - \frac{n'_s \omega_s}{c}\sqrt{1 - \left(\frac{\sin\theta_s}{n'_s}\right)^2} - \frac{n'_i(\omega_p - \omega_s)}{c}\sqrt{1 - \left(\frac{\omega_s n_i \sin\theta_s}{n_s n'_i(\omega_p - \omega_s)}\right)^2} \quad (6)$$

where $|\tau|$ is the biphoton transmission amplitude, which is linked to the absorption coefficient in the gap via the Beer–Lambert law $\alpha = -2\ln(|\tau|)/L'$, $\Delta\vec{k}'$ is the wave vector mismatch in the gap, $\vec{k}'_{p,s,i}$ are wave vectors in the gap, $n'_{p,s,i}$ are the refractive indices in the gap, and indices $p, s, i$ refer to pump, signal, and idler photons, respectively.

The first term in Eq. 5a determines the SPDC line shape from an individual crystal, while the second term determines the modulation function due to the interference, see Fig. 1b. The biphoton amplitude transmission $|\tau|$ defines the visibility of the quantum interference. The phase difference between the two biphotons $\Delta\varphi$ determines the periodicity of the interference pattern. Remarkably, the interference pattern, given by Eq. 5a, depends not only on the phase acquired by the signal photon itself, but also on the phases acquired by idler and pump photons, even though they have different wavelengths. If the gap between crystals is filled with some material, it is possible to infer its refractive index and the absorption coefficient at the frequency of the idler (IR) photon by measuring changes in the interference pattern for the signal (visible) photon. Direct detection of IR photons is not required. We note that the concept of the nonlinear interference also finds its applications in imaging[23,24] and metrology[25].

**Spectral selection free nonlinear spectroscopy**

In our earlier work, parameters of the medium were obtained from the wavelength-angular spectra, see Fig. 1b[13]. The spectrum was obtained using a spectrometer and a slit. However, the slit filters a narrow stripe from the full cone of the SPDC emission thus eliminating a large portion of the useful SPDC signal. Also, some part of the signal is inevitably lost due to inefficiencies of optical elements (gratings, mirrors) of the spectrograph. Let us now consider the case when the spectrograph and the slit are excluded:

a) **The spectrograph is excluded.** In this case, the interference pattern is determined by the integration of Eq. 5a over the frequency of signal photons:

$$|F_{int}(\theta_s)|^2 \propto \int sinc^2\left(\Delta k(\omega_s, \theta_s)\frac{L}{2}\right)(1 + |\tau| \cdot \cos(\Delta\varphi))\, d\omega_s \quad (7)$$

The integral is given by the projection of the interference pattern onto the angular axis, see Fig. 1c. The periodicity of the interference pattern depends on the phase difference $\Delta\varphi$, and the visibility depends on both $|\tau|$ and on $\Delta\varphi$. Note that in this case, the spectral bandwidth is defined by the natural linewidth of the SPDC.

b) **The slit is excluded.** In this case, the whole k-spectrum of the SPDC is detected by a wide field detector. The k-spectrum of the SPDC has a circularly symmetrical shape and it is modulated due to the nonlinear interference, see Fig. 1d. Capturing the whole k-spectrum allows us to read-out interference fringes across all the radial directions at once, which improves the signal-to-noise ratio.

Note that the visibility of the interference also depends on the spatial overlap of signal and idler beams[21,22,24,26,27]. The overlap is defined by the interplay of the size of the interaction region, given by the pump beam diameter $d$ and the gap length $L'$, and the scattering angle of down-converted photons $\theta_s$ [21,22]. The following condition should be fulfilled in order to achieve high visibility of the interference: $(2L + L')\tan\theta_s \ll d/2$, see inset in Fig. 2. We introduce an additional function

$V(\theta_s)$ into Eq. 7, which accounts for decrease of the visibility due to a non-perfect beam overlap. Then, Eq. 7 takes the form:

$$|F_{exp}(\theta_s)|^2 \propto y_0 + A \cdot \int sinc^2\left(\Delta k(\omega_s, \theta_s)\frac{L}{2}\right)(1 + |\tau| \cdot V(\theta_s) \cdot \cos(\Delta\varphi))\, d\omega_s, \qquad (8)$$

where $y_0$ is the background level, and $A$ is the amplitude of the interference pattern.

The case with the vacuum gap between the crystals serves as a convenient reference point with known $\Delta k'(\omega_s, \theta_s)$ and $|\tau|$ ($n'_{vac(p,s,i)} = 1$ and $|\tau|_{vac} = 1$). Measurements with the vacuum also allow to account for optical losses at frequencies of signal, idler and pump photons in nonlinear crystals. Fitting the experimental results with Eq. 8 allows us to determine $\Delta k$, $L'$, $y_0$, $A$ and $V(\theta_s)$.

Introducing a material between the crystals changes $|\tau|$ and $\Delta k'(\omega_s, \theta_s)$, see Eqs. 5b, 6, 8, while all other parameters remain the same. Assuming that $n'_p$, $n'_s$ are known, then $n'_i$ and $\alpha_i$ for idler photons can be inferred from fitting of the data with Eq. 8. Note, that separation of the idler photon from the pump and signal photons will allow measurements without prior knowledge of properties of the medium in visible range.

**Experimental results**

The obtained experimental interference patterns are shown in Fig. 3. Without the slit and the spectrograph, we observe the full 2-dimensional k-spectrum. Figs. 3a,b show experimental data for the gap with vacuum and with the carbon dioxide at pressure P = 7 Torr, respectively. These two cases correspond to the phase matching angle $\theta$ =50.5° and $\lambda_s$ = 608.4-611.3, $\lambda_i$ = 4100-4240 nm. In this wavelength range of idler photons the carbon dioxide gas has a comparatively low absorption but a strong change in the refractive index due to anomalous dispersion. That is the reason why interference fringes experience a significant phase shift. Averaging the data across azimuthal angles (as shown by the red arrow in Fig. 3a) produces a one-dimensional interference pattern with high signal-to-noise ratio, see Fig. 4. Experimental data and the numerical simulations for the vacuum gap and the gap filled with the carbon dioxide are shown in Fig. 4a,b, respectively. Data points are in a good agreement with the theory.

Experimentally measured refractive index and the absorption coefficient of the carbon dioxide are shown in Fig. 5. In our fitting procedure, we assume that the absorption line for the carbon dioxide is described by the Lorentz oscillator model. Experimental points for the refractive index are shown in Fig. 5a along with the theoretical curve. Theoretical values for the refractive index are calculated using Kramers-Kronig relations based on the absorption coefficient given by HITRAN database[28]. Spectral resolution of our measurements is given by the natural linewidth of the SPDC spectrum, and it is equal to ~80 cm$^{-1}$. Theory for the absorption coefficient is taken from the HITRAN database with the calculation step 0.01 cm$^{-1}$ and 80 cm$^{-1}$ rectangular function resolution.

We also measure the dependence of the refractive index and the absorption coefficient on the pressure, see Fig. 6. Theory is taken from the HITRAN database with the calculation step of 0.01 cm$^{-1}$ and rectangular function resolution of 80 cm$^{-1}$.

**DISCUSSION**

Measured dependencies of the refractive index and the absorption coefficient on the wavelength are in good agreement with the theory. In our earlier work, the resolution was 20 cm$^{-1}$ and it was given by the resolution of the spectrometer[13]. In contrast, the spectral resolution of the current method is 80 cm$^{-1}$ and it is limited by the natural linewidth of the SPDC. This limitation can be addressed by narrowing the SPDC spectrum using a longer SPDC crystal. For example, for a crystal length of $L$=5 mm the resolution reaches 15 cm$^{-1}$, which is close to the performance of the lab-based FTIR system. Using longer SPDC crystals will also increase the SPDC signal rate, which scales proportionally to the square of the coherence length[16], given by $L_{coh} = \frac{\sin(L\Delta k/2)}{\Delta k/2}$.

The experimental dependence of the refractive index and the absorption coefficient on the pressure matches the theory quite well. Some discrepancies occur due to the lack of sensitivity of our method at pressures less than 5 Torr. Also, we observe the mismatch with the theory at larger pressures due to the strong anomalous dispersion. The sensitivity of our method can be further improved by extending the distance between the crystal or/and by introducing a multipass configuration, for instance, by using Herriott cells.

In conclusion, we propose a method which allows simultaneous measurements of the refractive index and absorption coefficient of a medium in the IR range through measurements of the interference of visible photons. The method is based on nonlinear interference of entangled photons produced via SPDC from two nonlinear crystals. The key distinction of our method is that it excludes the need for the spatial and spectral selection. Our method yields high signal-to-noise ratio due to parallel readout of multiple spatial and frequency modes. It also makes the setup economical and compact. Accuracy of measurements of the refractive index is $5 \cdot 10^{-6}$; while the accuracy in measurement of the absorption coefficient is $0.05 \text{ cm}^{-1}$. The spectral resolution of the instrument in the IR range is $80 \text{ cm}^{-1}$ and it can be further improved by using longer crystals. We believe that this work contributes to practical appeals for implementation of the technique in material analysis and sensing applications.

**METHODS**

**Experimental realization**

Our experimental setup is shown in Fig. 2. A 532 nm continuous wave laser (30 mW power) is used as a pump. We substitute the scheme with two nonlinear crystals (discussed above) by a scheme consisting of a 1 mm MgO:LiNbO$_3$ crystal (doped with 5% Mg) and a silver mirror ($L'/2 = 10$ mm). The first biphoton is generated at the first passage of the pump beam throughout the crystal. Then all the three photons (pump, signal, and idler) are reflected back from the mirror with an additional phase shift π. The second biphoton is generated during the second pass of the pump through the crystal. The biphotons created at the first and second passes interfere. The coherence length of the laser is several tens of meters, which is significantly larger than the gap length $L'$.

The crystal is placed on a rotation stage and it can be tilted with the accuracy of 0.2°. Tilting the crystal provides the wavelength tunability of our system. The crystal is cut for a collinear Type-I ($e \rightarrow oo$) SPDC with the optical axis at $\theta = 50°$ to the surface. The crystal and the mirror are placed into a vacuum chamber with an optical window. The pump beam is reflected from the dichroic beam splitter DBS (Semrock, FF555-Di02-25x36) and directed into the vacuum chamber. The interference pattern of the signal photons is filtered from the pump and the idler photons by using a dichroic beam splitter and two notch filters (Semrock, NF03-532E-25). It is then focused by a lens with $f$=300 mm onto the sensor of a silicon CCD camera (Andor, iXon3 897). A bandpass filter BF (605±15 nm; Semrock FF01-605/15-25) is placed in front of the CCD camera to reduce the noise from the stray light.

The vacuum chamber is pumped down to low pressure (200 mTorr). The vacuum is used as the reference medium with $n'_{vac(p,s,i)} = 1$ and $|\tau|_{vac} = 1$. Then the carbon dioxide gas (purity 99.9%) is released into the chamber. The gas pressure is monitored by a gauge. We focus our study on a strong absorption line of the carbon dioxide at the wavelength of 4.27 μm (the corresponding signal wavelength is about 608 nm). The absorption coefficient and the refractive index are measured in the vicinity of this line.

We capture interference patterns by the CCD camera with 100 seconds acquisition time. Firstly, we collect the data for the vacuum gap at different crystal orientations. The background noise is acquired by tilting the crystal so that the SPDC signal does not pass through the bandpass filter.

**Numerical algorithm**

Before the fitting procedure, data processing is performed to eliminate the artifacts and reduce the speckle noise. The first step of the fitting algorithm is averaging the signal along the polar angles as shown by the arrow in Fig. 3a. The data is averaged across the whole k-spectrum of the SPDC (about 70 points per angle step), which leads to 8 times increase in the signal-to-noise ratio, compared to the case when the slit is used[9].

The experimental parameters $\Delta k$, $L'$, $y_0$, $A$, $V(\theta_s)$ are found using Eq. 8 by fitting the data for the interference fringes obtained with the vacuum gap.

Next, we fit the data obtained with the carbon dioxide gas. At this step parameters $\Delta k$, $L'$, $y_0$, $A$, $V(\theta_s)$ are fixed and parameters $\Delta k'(\omega_s, \theta_s)$, $|\tau|$ are unknown.

First, we evaluate $\Delta k'(\omega_s, \theta_s)$. Carbon dioxide refractive indices at the visible range $n'_{p,s}$ are assumed to be known, see the discussion above. Thus, in fact $\Delta k'(\omega_s, \theta_s)$ includes only one unknown parameter which is the refractive index at the idler photon wavelengths $n'_i$. Since the absorption line of the carbon dioxide is comparable with the SPDC linewidth, the algorithm takes into account the frequency dependence of the refractive index. We evaluate the refractive index according to the Lorentz model[29]:

$$n'_i = n'_{i0} - S \frac{(\omega - \omega_0)}{(\omega - \omega_0)^2 + (\Delta\omega/2)^2} \quad (9)$$

where $n'_{i0}$ is the refractive index at the IR region far from the resonance, $S = \frac{\pi N f e^2}{m \omega_0}$ ($N$ is the concentration, $f$ is the oscillator strength, $e$ is the electron charge, $m$ is the electron mass), $\omega_0$ is the resonance central frequency, $\Delta\omega$ is the resonance width.

To account for the pressure of the carbon dioxide we use partial refractive indices $n'_{m(p,s,i0)}$:

$$n'_{p,s,i0} = n'_{eff(p,s,i)} = \left(1 - \frac{P_{CO2}}{P_{atm}}\right) n'_{vac(p,s,i)} + \frac{P_{CO2}}{P_{atm}} n'_{CO2(p,s,i)} \quad (10)$$

where $P_{atm}$ is the atmospheric pressure, $P_{CO2}$ is the carbon dioxide pressure, $n'_{CO2(p,s,i)}$ is the carbon dioxide refractive index at the atmospheric pressure, which are known from a corresponding database[30]. By evaluating parameters $S$, $\omega_0$, $\Delta\omega$ we find the fit of data points by Eq. 8, which defines the refractive index $n'_i$.

At the last step, we evaluate the absorption coefficient, which is given by $\alpha_i = -2\text{Ln}(|\tau|)/L'$. Since the carbon dioxide absorption linewidth is comparable with the idler photon spectrum, the visibility of the interference pattern decreases with the injection of carbon dioxide. Our algorithm finds the optimal fit of the data by Eq. 8, and outputs parameters: $S$, $\omega_0$, $\Delta\omega$, $\alpha_i$.

## Acknowledgements

This work is supported by NRF CRP grant NRF—CRP14-2014-04.

## Author contributions

A.V.P. and S. L. built the experiment setup and conducted measurements, A.V.P. carried out theoretical analysis, S.L. developed numerical algorithms, D.A.K. participated to the building of the experimental setup, L.A.K. conceived the idea of the experiment and supervised the work. All the authors contributed to writing the manuscript.

## Additional information

**Competing financial interests:** The authors declare no conflict of financial interests.

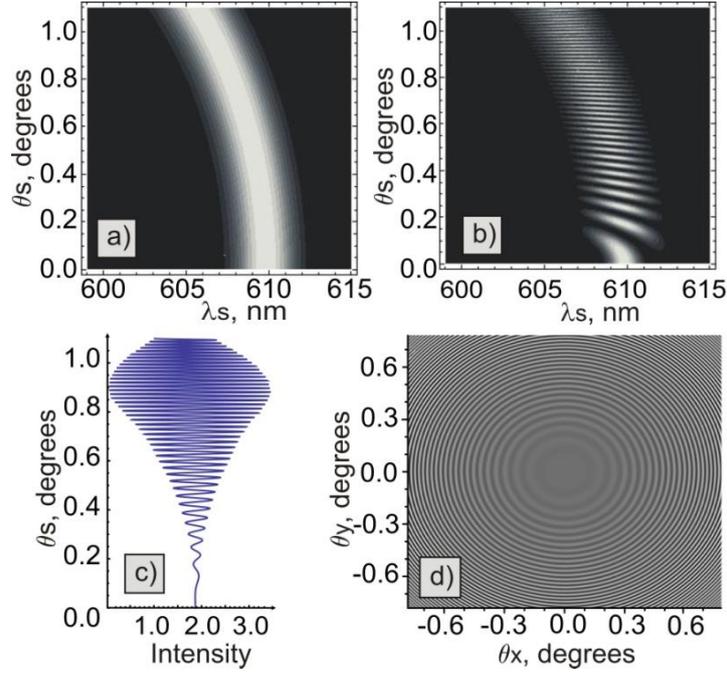

**Figure 1.** (**a**) Theoretical frequency-angular spectrum of the signal photon of the SPDC from one LiNbO3 crystal (crystal length L=1 mm, phase matching angle θ = 50.5°, $\lambda_p$ = 532 nm, $\lambda_i$ = 4100-4240 nm); (**b**) interference pattern for the two LiNbO3 crystals (same as in (a)) at a distance of L′ = 20 mm separated by a vacuum gap; (**c**) projection of the two-dimensional interference pattern shown in b) onto the angular axis; (**d**) corresponding interference fringes in a two-dimensional k-space.

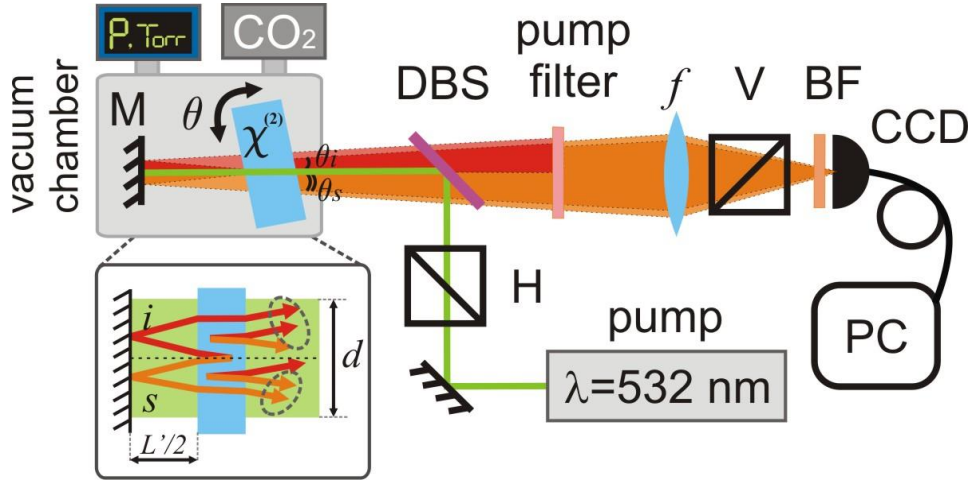

**Figure 2.** Experimental setup. A CW laser at 532 nm is reflected by a dichroic beamsplitter (DBS) and pumps a MgO:LiNbO$_3$ crystal, where SPDC occurs. The interferometer is formed by a crystal and a mirror, which are placed inside a vacuum chamber, where the carbon dioxide gas is injected. The crystal is on a rotation stage, for controlling the wavelength of the SPDC. Scattering angles of the signal and idler photons are indicated by $\theta_s$ and $\theta_i$, respectively. The interference pattern is directly imaged by a CCD camera placed at the focal plane of the lens *f*. The signal is filtered by a dichroic beamsplitter (DBS), a band pass filter (BF), and a polarizer (V). Idler and signal photons are indicated by red and orange colors, respectively (overlapped beams are shown by brighter colors). Inset shows trajectories of beams in the interferometer, where *d* is the pump beam diameter and $L'$ is the gap full length.

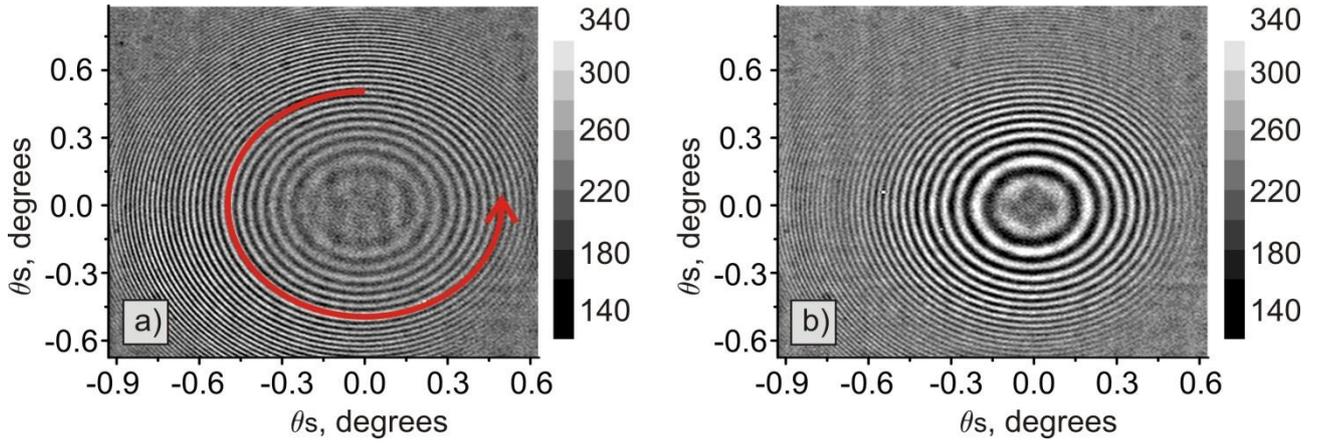

**Figure 3.** Experimental interference patterns for SPDC from one crystal and a mirror at a distance of $L'/2 = 10$ mm with $\theta = 50.5°$ phase matching angle, $\lambda_s = 608.4\text{-}611.3$ nm, $\lambda_i = 4100\text{-}4240$ nm wavelength for the signal/idler waves. (**a**) The pattern for the vacuum gap; (**b**) the pattern for the gap filled with the carbon dioxide gas at 7 Torr pressure. The red arrow in (a) depicts the averaging over azimuthal angles.

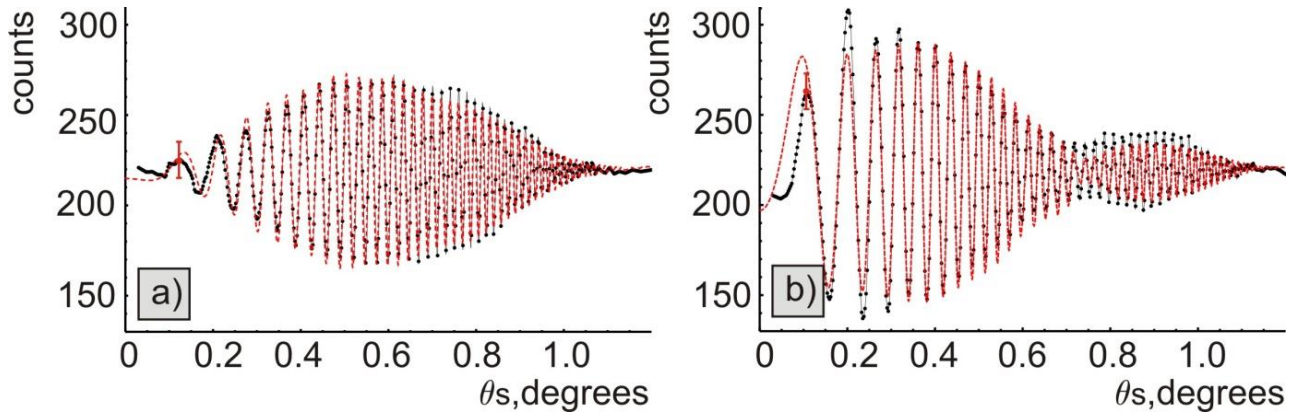

**Figure 4.** One dimensional interference patterns obtained by averaging of experimental data in Fig.2 over azimuthal angles. (**a**) Interference pattern for the vacuum gap; (**b**) interference pattern for the gap filled with the carbon dioxide gas at 7 Torr pressure. Experimental points are indicated by black dots, fitted curves are shown by red dashed lines. Typical standard deviation of the data is indicated by red bars.

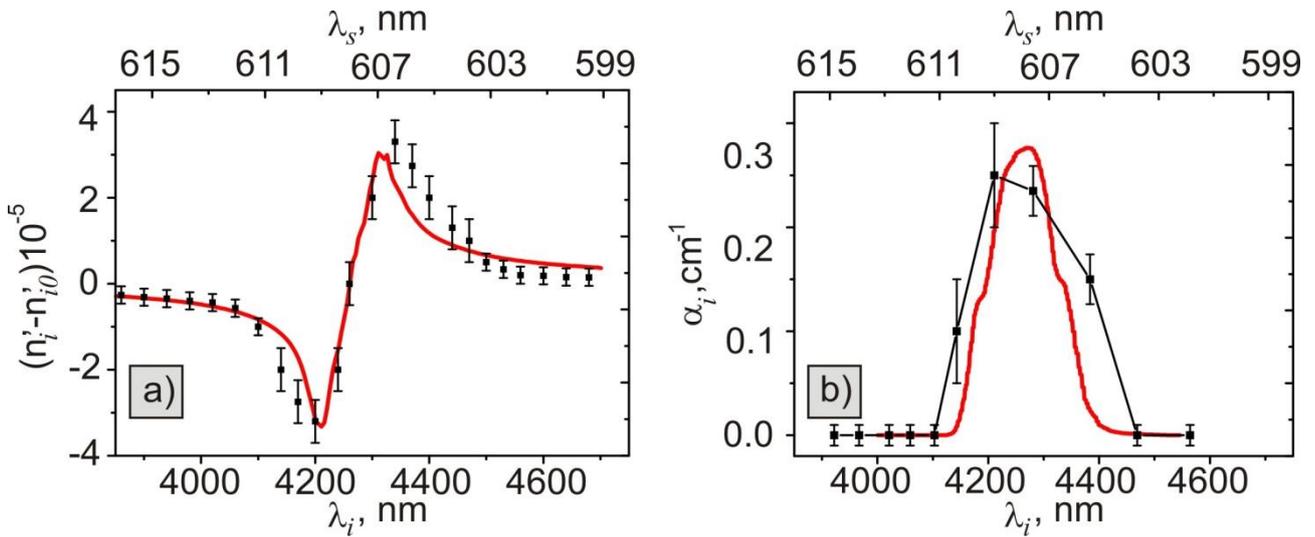

**Figure 5.** Measured values for the (**a**) refractive index and (**b**) absorption coefficient of the carbon dioxide gas at P=7 Torr pressure. Black dots show calculated values, red lines show theoretical curves. Bottom horizontal axis shows actual wavelength at the IR (for idler photons), while the top horizontal axis indicates corresponding wavelengths of the measured signal photon.

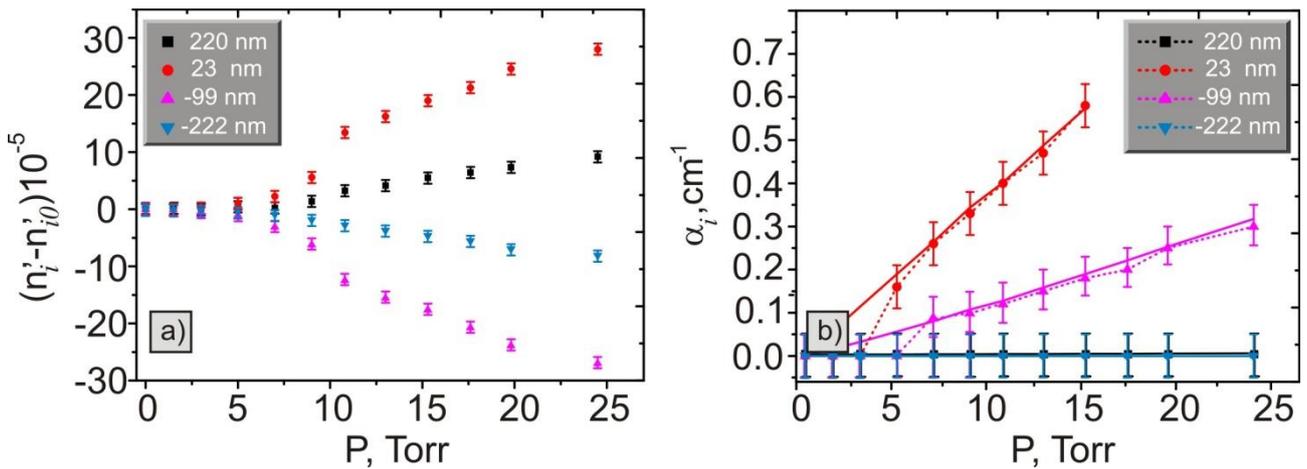

**Figure 6.** Pressure dependence of the (**a**) refractive index and (**b**) absorption coefficient at different wavelengths. The legend shows the detuning from the resonant wavelength $\lambda_0 = 4266$ nm. The theory for the absorption coefficient is shown by solid lines. Note, that the data points in Fig. 6b for both 220 nm and -222 nm detuning almost coincide due to the low value of the absorption coefficient.